\documentclass{article}
\usepackage{./epsfig}
\usepackage{./sprocl}
\usepackage{./mcite}

\def\lsim{\mathrel{\rlap{\lower2pt\hbox{\hskip1pt\small$\sim$}}
    \raise2pt\hbox{\small$<$}}} 
\def\gsim{\mathrel{\rlap{\lower3pt\hbox{\hskip0pt\small$\sim$}}
    \raise2pt\hbox{\small$>$}}} 

\newcommand{\D}{{\rm d}}

\newcommand{\mev}{{\rm Me}\kern-1.pt{\rm V}}
\newcommand{\gev}{{\rm Ge}\kern-1.pt{\rm V}}
\newcommand{\gevsq}{\mbox{$\mathrm{{\rm Ge}\kern-1.pt{\rm V}}^2$}}

\newcommand{\rhoz}{\mbox{$\rho^0$}}
\newcommand{\jpsi}{\mbox{$J/\psi$}}

\newcommand{\qsq}{\mbox{$Q^2$}}

\newcommand{\phih}{\mbox{${\phi}_h$}}
\newcommand{\Phih}{\mbox{${\Phi}_h$}}

\newcommand{\thetah}{\mbox{${\theta}_h$}}

\newcommand{\abst}{\mbox{$|t|$}}
\begin{document}

\vspace*{-2.5cm}
\mbox{ }\hfill {\mbox{DESY 99-163}}\\
\mbox{ }\hfill {\mbox{BONN-HE-99-05}}\\
\mbox {}\hfill {\mbox{October, 1999}}\\*[9mm]
\title{
Topical Results on Vector-Meson Production\\ from the HERA Collider
Experiments\footnote{Talk presented at the Joint INT/Jefferson Lab
Workshop on Exclusive and Semi-Exclusive Processes at High Momentum
Transfer, 20-22 May 1999, Jefferson Lab, Newport News, Virginia, USA}
}

\author{James A. Crittenden\\for the H1 and ZEUS collaborations}

\address{Physikalisches Institut, University of Bonn, Nu{\ss}allee 12, 53115 Bonn, Germany}


\maketitle\abstracts{
The HERA collider experiments H1 and ZEUS 
have established extensive
measurement programs for diffractive vector-meson production processes
during the first six years of their operation.
The results provide stringent phenomenological tests of 
quantum chromodynamical descriptions of hard diffraction.
We discuss recent topical results on $\Upsilon$ photoproduction, on
decay-angle analyses of {\rhoz}, {$\phi$} and {\jpsi} electroproduction and 
on {\rhoz} and {$\phi$} photoproduction at high momentum transfer.
}
\section{Introduction}
The HERA collider experiments H1 and ZEUS have developed measurement
programs of diffractive vector-meson production in electron-proton
collisions, extending the rich history of such investigations
in the 1960s and 1970s to the modern era. Many of the topics addressed
by the data have as sole precedent investigations performed before the
development of quantum chromodynamics as the fundamental theory
of the strong interaction. As a result, recent years have witnessed
rapid expansion in a field of study in which the predictions of 
an asymptotically free field theory are tested for the first time
in new domains of energy and photon virtuality.
In light of the recent impressive theoretical successes in
describing hard diffractive phenomena, in particular the factorization
theorems proven for exclusive~\cite{pr_56_2982} and semi-exclusive~\cite{pr_53_3564,*pr_54_5523,*pl_449_306} vector-meson production, 
these new measurements are providing an extensive phenomenology
useful in testing the precepts of quantum chromodynamics in their
application to diffractive processes.

In this report, we discuss three topics of particular interest to this
workshop: the first measurements of $\Upsilon$ photoproduction, the results
on the helicity structure of {\rhoz}, $\phi$ and {\jpsi} electroproduction
and the semi-exclusive photoproduction of {\rhoz} and $\phi$ mesons at
high momentum transfer.
\section{{$\Upsilon$} Photoproduction}
\begin{figure}[htbp]
\begin{center}
\epsfig{file=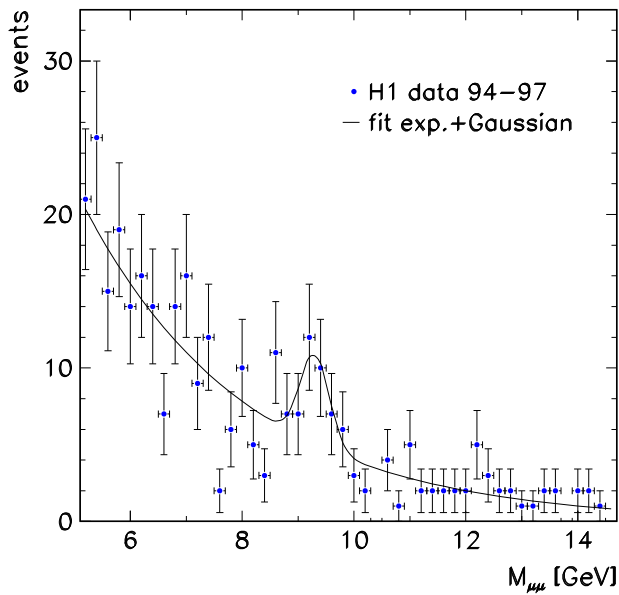, width=0.68\textwidth, bbllx=20pt, bblly=10pt, bburx=512pt, bbury=485pt, clip=}
\caption{
\label{fig:h1upsilon} 
Invariant mass spectrum for diffractively 
photoproduced muon pairs reported by the H1
collaboration.\hspace*{-0.5ex}\protect\cite{ichep98_574} The curve shows the result of
a fit to a sum of exponential and Gaussian functions
}
\end{center}
\end{figure}

\begin{figure}[htbp]
\begin{center}
\epsfig{file=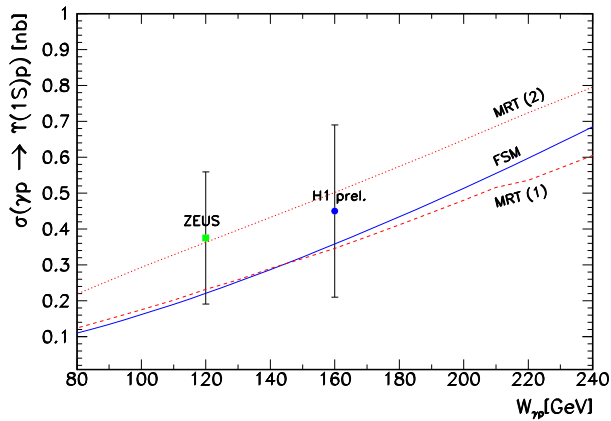, width=0.68\textwidth, bbllx=16pt, bblly=25pt, bburx=528pt, bbury=377pt, clip=}
\caption{
\label{fig:upsrot} 
Measurements from the H1 and ZEUS collaborations of the elastic $\Upsilon$(1S)
photoproduction cross section. The error bars show the quadratic sum of
statistical and systematic uncertainties. The curves show the results of 
QCD-based calculations which take into account a variety of effects beyond leading order~\protect\cite{jhep99_02_002,*pl_454_339}}
\end{center}
\end{figure}

In stark contrast to the first published observation of 
photoproduction of the {\jpsi} meson 
a few months following its discovery in 1974, 
twenty years passed between the discovery of the $\Upsilon$ meson
and the first report of its production in photon-proton interactions
by the ZEUS experiment in 1997.\hspace*{-1.ex}~\cite{pl_437_432} This 
measurement of 
exclusive $\Upsilon$ 
photoproduction at a photon-proton center-of-mass energy, $W$, 
of 120~\gev, has since been confirmed by the H1 collaboration, who have
released a preliminary result~\cite{ichep98_574} for the cross section at
$W$=160~\gev. Each
experiment identifies  the $\Upsilon$ family via a resonant structure in
the observed dimuon mass spectrum, which is dominated by the the
continuum background from the Bethe-Heitler pair-production process.
Figure~\ref{fig:h1upsilon} shows the preliminary measurement of this
mass spectrum reported by H1. The H1 spectrum shows a peak
which leads to an estimated signal of \mbox{$8.3\pm3.9$} events.
The mass resolution provided by the
central tracking chambers and magnetic field does not allow the separation
of the $\Upsilon, \Upsilon^\prime$ and $\Upsilon^{\prime\prime}$ in
either of the experiments, requiring the use of estimates of
their relative production in order to extract a production cross section
for the $\Upsilon$ alone. The resulting determinations of the cross section
for the exclusive process   
$\gamma p \rightarrow \Upsilon{\rm (1S)} p$, shown in  Fig.~\ref{fig:upsrot},
exceed the 
leading-order QCD estimate~\cite{hep99_05_288} by a factor of 5-10.
Detailed theoretical investigations~\cite{jhep99_02_002,*pl_454_339} into 
reasons for the large cross
section have emphasized the importance of the evolution of the parton
density functions (which are necessarily skewed by the
kinematic lower bound on the momentum transferred to the proton resulting
from the rest mass of the $\Upsilon$(1S)) and that of the contribution
from the real part of the amplitude.

These initial measurements establishing the large cross section 
were based on an integrated luminosity of about
40~pb$^{-1}$ for each experiment. 
The HERA luminosity upgrade program and operation schedule
through 2005 will permit an increase in the $\Upsilon$ candidate event sample
of at least a factor of twenty and 
is likely to enable decay-angle studies  of the type
described below for the lighter vector mesons.
\section{Helicity Analyses of {\rhoz}, {$\phi$} and {\jpsi} Electroproduction}
The simplicity of the final state in the exclusive production of
vector mesons permits
detailed investigations of the decay-angle distributions, which have provided
much
 information on the helicity structure of the diffractive production
mechanism. 
These analyses of the decay-angle
distributions have been employed as a means of measuring  the
ratio of the cross sections for longitudinal and transverse photons, $R$. 
This type of study provided another early success of the perturbative QCD models
of exclusive diffractive processes by verifying the prediction that the longitudinal
cross section exceed the transverse cross section at high {\qsq}\hspace*{-1.2ex}.

Three angles suffice to completely describe the exclusive electroproduction
of vector mesons, as shown in Fig.~\ref{fig:defangles} for the example of 
{\rhoz} production: the azimuthal angle between the scattering plane and the production plane, 
$\Phi_h$, and the two $\rho^0$ decay angles: 
\begin{figure}[htbp]
\begin{center}
\epsfig{file=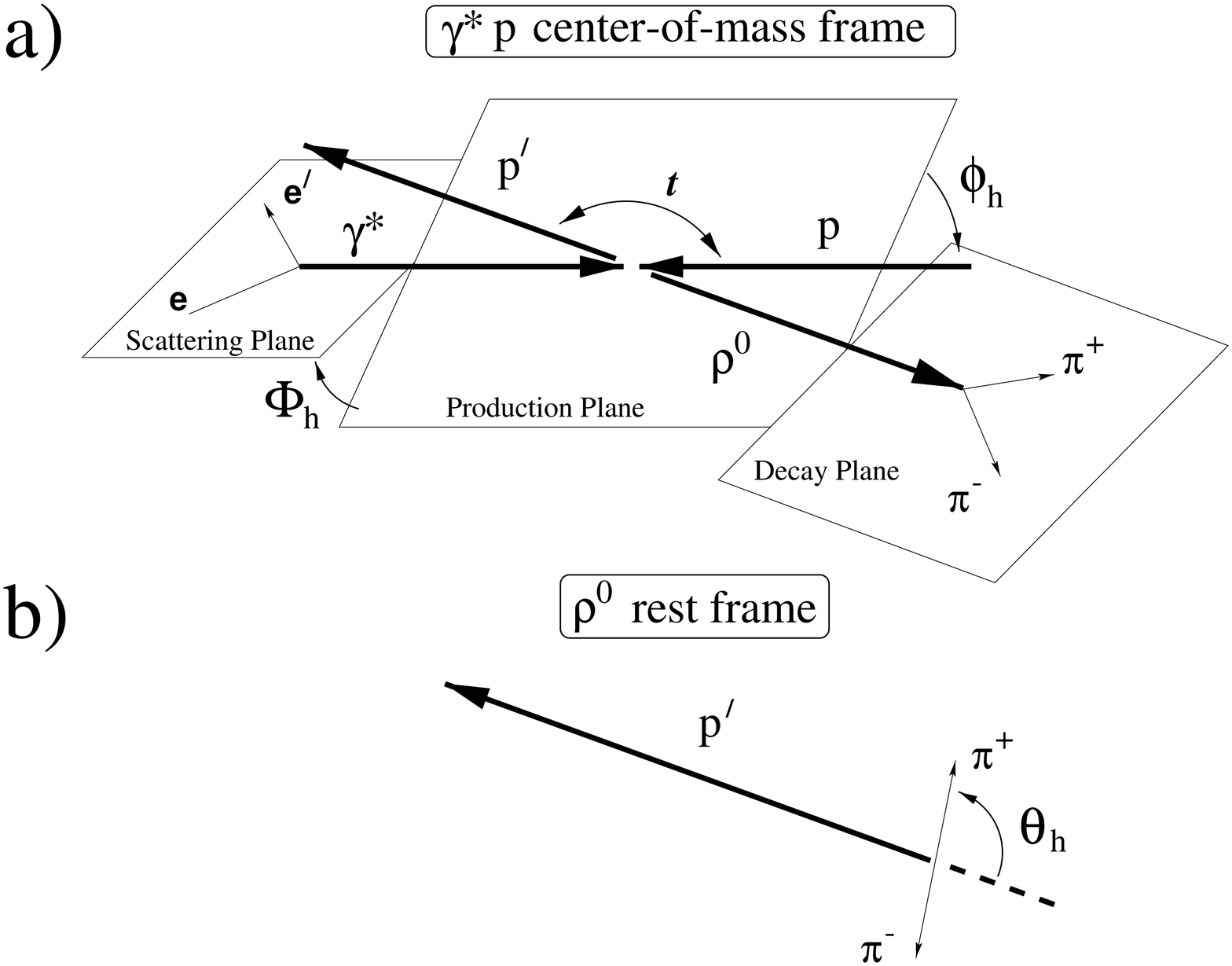, width=0.68\textwidth}
\caption{
\label{fig:defangles} 
Schematic diagram of the process $ep \rightarrow e \rho^0 p$ (a) in the
virtual-photon/proton center-of-mass system, and (b) in the rest frame of
the $\rho^0$ meson\vspace*{-2mm}
}
\end{center}
\end{figure}
$\phi_h$, the azimuthal angle 
between the production and decay planes, defined in either 
the virtual-photon/proton system or in
the $\rho^0$ rest frame, and $\theta_h$, which is the polar angle of the 
positively charged decay product, defined with respect to the direction
of the $\rho^0$ momentum vector in the virtual-photon/proton system, which is 
opposite to the momentum vector of the final-state
proton in the rest frame of the $\rho^0$ meson. This latter choice of 
spin-quantization axis defines the helicity frame, in which helicity
conservation was found to hold approximately in exclusive {\rhoz} photoproduction
experiments at SLAC in the early 1970s.\hspace*{-0.7ex}~\cite{pr_5_545,pr_7_3150}  

\begin{figure}[htbp]
\begin{minipage}[t]{0.48\textwidth}
\begin{center}
\epsfig{file=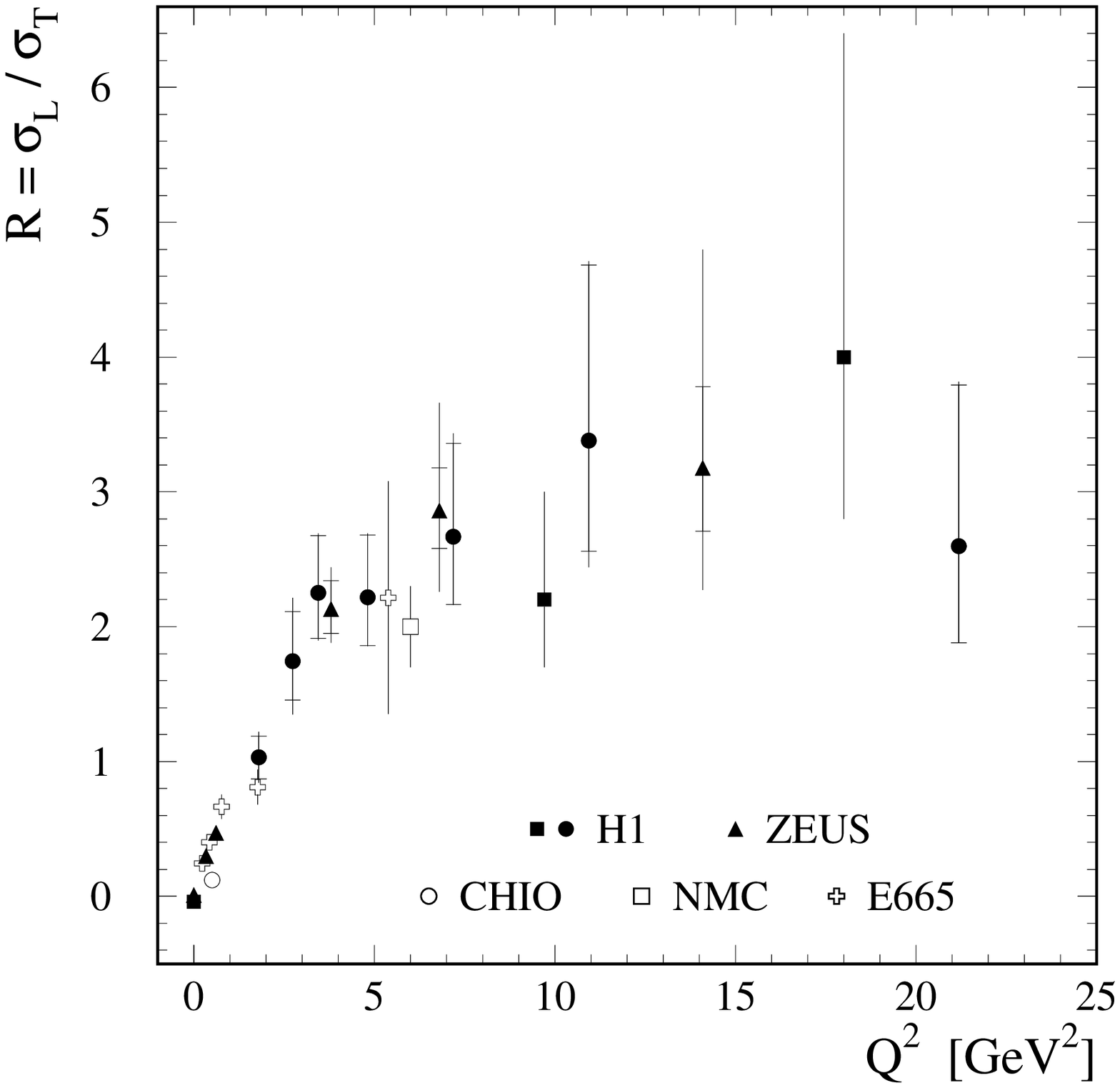, width=\textwidth}
\caption{ 
\label{fig:h1_R}
Measurements of the ratio of exclusive {\rhoz} electroproduction cross sections
for longitudinal and transverse photons, $R$, as a function of photon
virtuality, {\qsq}
} 
\end{center}
\end{minipage}
\hfill
\begin{minipage}[t]{0.48\textwidth}
\begin{center}
\epsfig{file=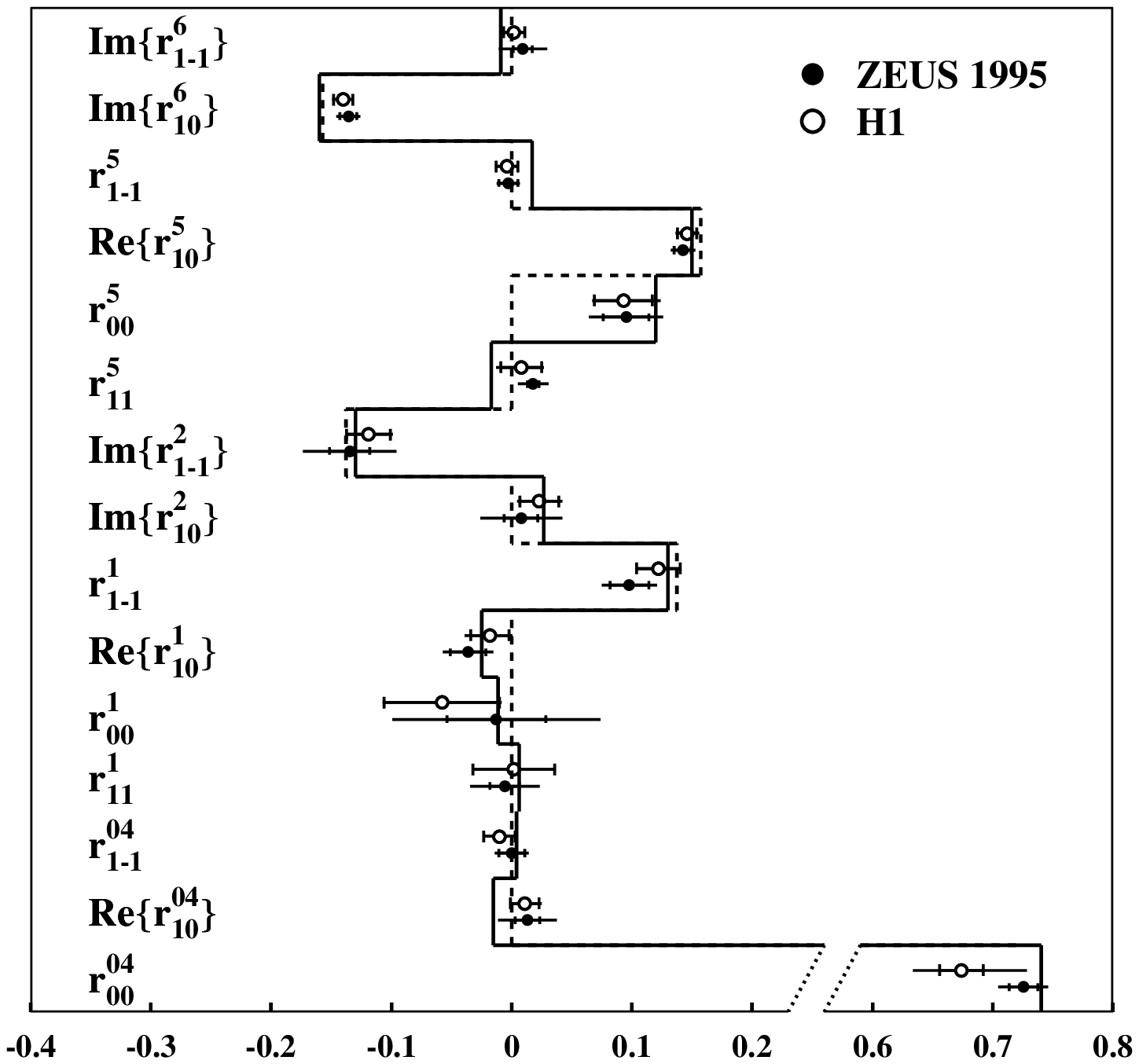, width=\textwidth, bbllx=43pt, bblly=37pt,
  bburx=462pt, bbury=426pt, clip=}
\caption{ 
\label{fig:me_dis}
Combinations of spin-density matrix elements measured for {\rhoz}
electroproduction by the H1~\protect\cite{dr_99_010} and ZEUS~\protect\cite{dr_99_102} collaborations.
See text for full description
} 
\end{center}
\end{minipage}
\end{figure}

Following the work of Schilling and Wolf,\hspace*{-0.7ex}~\cite{np_61_381} 
the three-dimensional angular distribution 
for this decay of a spin-1 state to two spinless particles
has been parameterized as follows:
\begin{eqnarray}
W({\cos\theta_h},\phi_h,\Phi_h) &\hspace*{-1pt}=&\hspace*{-1pt} \frac{3}{4\pi} \biggl[ 
\frac{1}{2}(1-r^{04}_{00})+\frac{1}{2}(3r^{04}_{00}-1)\cos^2{\theta_h} \nonumber\\
& &\hspace*{-26pt}-\sqrt{2}~{\rm Re}\{r^{04}_{10}\}\sin{2\theta_h}\cos{\phi_h} -r^{04}_{1-1}\sin^2{\theta_h}\cos{2\phi_h} \nonumber\\
& &\hspace*{-26pt}-\epsilon\cos{2\Phi_h}(r^1_{11}\sin^2{\theta_h}+r^1_{00}\cos^2{\theta_h}-\sqrt{2}~{\rm Re}\{r^1_{10}\}\sin{2\theta_h}\cos{\phi_h}\nonumber\\
& &\hspace*{-26pt}\hspace*{1.8cm}-r^1_{1-1}\sin^2{\theta_h}\cos{2\phi_h}) \nonumber\\
& &\hspace*{-26pt}-\epsilon\sin{2\Phi_h}(\sqrt{2}~{\rm Im}\{r^2_{10}\}\sin{2\theta_h}\sin{\phi_h}+{\rm Im}\{r^2_{1-1}\}\sin^2{\theta_h}\sin{2\phi_h})\nonumber\\
& &\hspace*{-26pt} +\sqrt{2\epsilon (1+\epsilon)}\cos{\Phi_h} (r^5_{11}\sin^2{\theta_h}+r^5_{00}\cos^2{\theta_h} \nonumber\\
& &\hspace*{-26pt}\hspace*{2.cm}-\sqrt{2}~{\rm Re}\{r^5_{10}\}\sin{2\theta_h}\cos{\phi_h} -r^5_{1-1}\sin^2{\theta_h}\cos{2\phi_h})\nonumber\\
& &\hspace*{-26pt} +\sqrt{2\epsilon (1+\epsilon)}\sin{\Phi_h} (\sqrt{2}~{\rm Im}\{r^6_{10}\}\sin{2\theta_h}\sin{\phi_h} \nonumber\\
& &\hspace*{-26pt}\hspace*{3.7cm}+{\rm Im}\{r^6_{1-1}\}\sin^2{\theta_h}\sin{2\phi_h}) \biggr],
\label{full_equation}
\end{eqnarray}
where the superscripts of the combinations of spin-density matrix elements
correspond to the helicity degrees of freedom of the virtual photon, and
the subscripts to those of the dipion state. 
The fifteen coefficients   $r^{04}_{ik}$,
$r^{\alpha}_{ik}$ are 
related directly to various combinations of
the helicity amplitudes, $T_{\lambda_{\rho} \lambda_{\gamma}}$, where 
$\lambda_{\rho}$ and $\lambda_{\gamma}$ are the helicities of the $\rho^0$
meson and of the photon, respectively. The assumption that helicity is
conserved in the photon/vector-meson transition when the amplitudes
are defined in the helicity frame ({``s-channel helicity conservation''} or 
``SCHC''),
with the consequence that the degree of {\rhoz} polarization is equal
to the ratio of the longitudinal and transverse cross sections, allows the
extraction of this ratio $R$ from the distribution in polar angle alone. A compilation
of determinations of $R$ via this method is shown in Fig.~\ref{fig:h1_R}. 
The data from the fixed-target muon experiment at FNAL, E665,\hspace*{-0.7ex}~\cite{zfp_74_237}
and the low-{\qsq} data from the ZEUS collaboration identify a region
of transition to increasing values of $R$, with the longitudinal cross section becoming
dominant for $\qsq\gsim 2~\gevsq$\hspace*{-1.2ex}.\hspace*{1.2ex} 
Together with the latest results from the H1 collaboration,\hspace*{-1.2ex}~\cite{dr_99_010}
these data indicate that the value of $R$ reaches a plateau for \mbox{$\qsq\gsim 5~\gevsq$}.

The H1~\cite{dr_99_010} and ZEUS~\cite{dr_99_102} collaborations
have recently completed analyses
of the three-dimensional angular distribution for data samples of a few
thousand events, extracting the fifteen
coefficients $r^{04}_{ik}$, $r^{\alpha}_{ik}$. Figure~\ref{fig:me_dis}
shows the results from the ZEUS collaboration 
in the kinematic region \mbox{$3<{\qsq}<30~\gevsq$}\hspace*{-1.2ex},
\mbox{$40 < W <120~\gev$} and \mbox{${\abst}<0.6~\gevsq$}\hspace*{-1.2ex}, 
comparing them to the results
from the H1
collaboration in a similar kinematic region, 
and to a calculation by
Ivanov and Kirschner~\protect\cite{pr_58_114026} (solid line). 
The inner error bars represent the statistical
uncertainty; the outer error bars show the quadratic sum of
statistical and systematic uncertainties. 
Also shown
are the values of the coefficients predicted according to the hypothesis 
of helicity
conservation in the $s$-channel amplitudes~\cite{np_61_381} (dashed line).
Of particular interest is the violation of SCHC evinced in the non-zero value
for the coefficient
$r^5_{00}$, which is manifested 
in the distribution in the azimuthal angle between the
scattering and production planes ({\Phih}). This observation implies
a small non-zero single-flip amplitude for the production of {\rhoz} mesons in 
helicity state 0 from transverse photons. Such a violation has now been
reproduced in two further recent calculations based on differing
assumptions.\hspace*{-0.7ex}~\cite{schnc_dis99} These measurements show the level of
violation of SCHC to be small enough that the effect of the assumption of
SCHC in the earlier determinations of $R$ is much smaller than the other
sources of uncertainty in those determinations.

The H1 collaboration has recently extended their investigation to the
proton-dissociative electroproduction of {\rhoz} mesons,\hspace*{-1.0ex}~\cite{eps99_157n}
verifying with similar significance that the violation
of SCHC observed for the
exclusive case holds also when the proton dissociates.

Preliminary results of a full decay-angle analysis 
with similar statistical power for exclusive $\phi$ electroproduction
in the same kinematic region
have been presented by the ZEUS collaboration last year.\hspace*{-1.0ex}~\cite{ichep98_7923a}
They find values for the fifteen combinations of matrix elements consistent
with those found for the {\rhoz} meson, exhibiting a similar helicity
violation and comparable values for $R$.

The ZEUS collaboration also performed the 
analysis~\cite{dr_99_102}  in the low-{\qsq} region \mbox{$0.25<Q^2<0.85$~GeV$^2$}, finding
a value for $r^5_{00}$ of similar magnitude, but also distinct indications
for a more complex pattern of helicity violation, including double-flip
contributions, as previously found in {\rhoz} photoproduction (see Sect.~\ref{sec:hight}).

Statistical limitations have precluded such full decay-angle analyses
for {\jpsi} electroproduction; however, both the H1~\cite{dr_99_010} and ZEUS~\cite{epj_6_603}
collaborations have published measurements of $R$ for the exclusive
electroproduction of {\jpsi} mesons under the assumption of SCHC. 
The values are found to be significantly
smaller than those found for the {\rhoz} at similar {\qsq}\hspace*{-1.2ex}.

\section{Photoproduction of {\rhoz}, {$\phi$} Mesons at High Momentum Transfer}
\label{sec:hight}
The  
photoproduction of light vector mesons with high transverse momenta
has raised interest recently~\cite{pl_375_301,pr_53_3564,*pr_54_5523,*pl_449_306} as a means of investigating the r\^ole of 
the momentum transferred to the proton
in establishing the scale of the interaction,
since the photoproduction of {\rhoz} mesons at low {\abst} has been
shown to be governed by a soft diffractive production mechanism.
However, diffractive photoproduction 
of the light vector 
mesons has been shown to be dominated by the proton-dissociative process
for squared momentum transfers ${\abst}\gsim 0.5~\gevsq$\hspace*{-1.5ex},\hspace*{1.2ex} well
below the perturbative regime.\hspace*{-0.7ex}~\cite{ichep98_788} The investigation of exclusive vector-meson
production at high {\abst} requires a trigger on the elastically
scattered proton and such studies at HERA have lacked the integrated
luminosity necessary to measure the small cross sections at such high
values of {\abst}.
However, the ZEUS collaboration~\cite{eps99_499} has recently
employed a photoproduction-tagging method to investigate the
proton-dissociative production of {\rhoz} mesons for values of {\abst} up to
11~{\gevsq}\hspace*{-1.2ex}.\hspace*{1.2ex} The trigger conditions
required the scattered positron to be detected in a special-purpose
tungsten/scintillator calorimeter located 3~cm from the positron beam axis, 44~meters distant
from the nominal e$^+$p interaction point in the positron-beam flight direction.
The position of this photoproduction tagger determines the accepted range
of energy
lost by the positron to the photon which interacts with the proton, thus
restricting the $W$ range to the region $80<W<120~\gev$.
Since the transverse momentum of the final-state positron is thus required to
be small (${\qsq}<0.01~{\gevsq}$), the transverse momentum of the
{\rhoz} ($p_{\rm t}$) detected in the central detector via its dipion
decay provides an accurate approximation
for the square of the four-momentum transferred to the proton: $t\simeq
-p^2_{\rm t}$. 
Offline data-selection criteria include the reconstruction of exactly two
tracks from the interaction vertex and reject events with calorimetric energy
deposits in the rear and barrel sections of the calorimeter which are not
associated with the extrapolation of either track. The selected events exhibit
a substantial rapidity
gap between the dissociated nucleonic 
system and the two tracks. Even
for the highest values of {\abst}, the decay pions are in the rear half of the central detector.

\begin{figure}[ht]
 \vspace*{-3mm}
\begin{center}
\epsfig{file=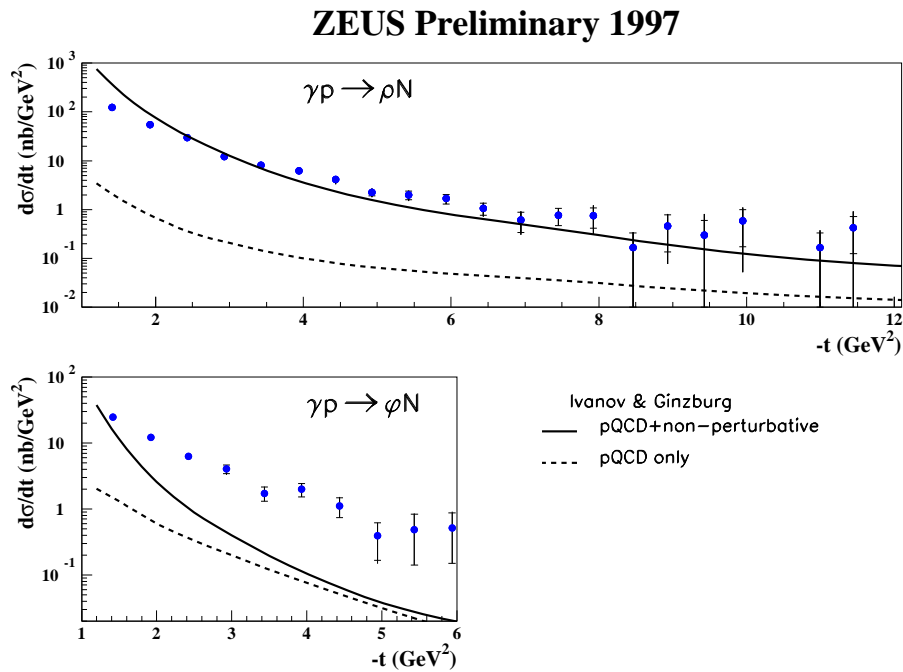, width=\textwidth, bbllx=18pt, bblly=228pt,
  bburx=540pt, bbury=618pt, clip=}
\vskip -3mm
\caption{
\label{fig:rho}
Preliminary measurements
of the differential cross sections $\frac{\D\sigma}{\D t}$ for
the proton-dissociative photoproduction of {\rhoz} and $\phi$ mesons in the energy range \mbox{$80 < W
  <120~\gev$} recently presented by the ZEUS collaboration~\protect\cite{eps99_499}\hspace*{-2.2ex}.\hspace*{2.2ex} The inner error bars represent the statistical uncertainties;
the outer error bars show the quadratic sum of statistical and systematic uncertainties. 
The results of a QCD calculation by Ivanov and
Ginzburg~\protect\cite{pr_53_3564,*pr_54_5523} (solid line) which
distinguishes the purely perturbative contribution (dashed line) are shown for
comparison
}
\end{center}
\end{figure}

Figure~\ref{fig:rho} shows the 
differential cross sections, $\frac{\D\sigma}{\D t}$, measured for
{\rhoz} and $\phi$ mesons. These exhibit extremely hard spectra;
a fit to the form $\frac{\D\sigma}{\D t} \propto (-t)^{-n}$
results in a value for $n$ of approximately~3.
The results are compared to a QCD calculation by Ivanov
and Ginzburg~\cite{pr_53_3564,*pr_54_5523} which estimates both perturbative
and non-perturbative contributions. In this model, the non-perturbative
contribution is found to account for the hardness of the spectrum and
to dominate in the region covered by the measurements
for the {\rhoz} meson. 
The calculation underestimates the $\phi$ cross section over most of the $t$ range covered. 
The comparison of the data to the calculation of the
purely perturbative contribution alone shows a good description of the shape of the $t$
dependence, whereas the magnitude of the cross section is underestimated by more
than one order of magnitude. 
The model thus fails to reproduce the ratio of $\phi$ production to that for the {\rhoz}
meson at high {\abst}, which is measured to be consistent with
the value of 2/9 expected for a flavor-independent production mechanism.

The ZEUS collaboration has employed this 
photoproduction-tagging technique to
perform decay-angle
analyses of the  diffractive
photoproduction~\cite{hep99_06_005,eps99_499} of pion pairs in the {\rhoz}
mass
region at values of {\abst} up to 
4~{\gevsq}\hspace*{-1.2ex}.\hspace*{1.2ex} In this study, the small contribution from the elastic process was
not subtracted. The decay-angle distribution was parameterized in terms of combinations
of spin-density matrix elements in the Schilling-Wolf convention,\hspace*{-1.0ex}~\cite{np_61_381}
$r^{04}_{ij}$, as
\begin{eqnarray}
\begin{array}{ll}{W(\thetah,\phih)}={3\over 4\pi} \biggl[
\frac{1}{2} \left(1-{{r^{04}_{00}}}\right)+\frac{1}{2} 
\left(3{{r^{04}_{00}}}-1\right)\cos^2{\thetah} \\[4mm]
-\sqrt{2}{{{\rm Re}(r^{04}_{10})}} \sin 2\thetah\cos\phih
-{{r^{04}_{1-1}}}\sin^2\thetah\cos{2\phih}
\;\biggr],\end{array}
\end{eqnarray}
where the three-dimensional distribution (Eq.~\ref{full_equation}) has been
averaged over the unmeasured azimuthal angle between the positron scattering
plane and the {\rhoz} production plane, and thus no longer distinguishes the
photon helicity states~$\pm 1$.

Under the assumption that the dipion final state is produced with one
unit of angular momentum, and neglecting the contribution by longitudinal
photons, these combinations of matrix elements are
related to the helicity amplitudes, \mbox{$T_{\lambda_{\rho} \lambda_{\gamma}}$}, as follows~\cite{np_61_381,dr_99_102}:
\begin{eqnarray}
r_{00}^{04} \simeq \frac{\strut |T_{01}|^2}{\strut
    |T_{01}|^2+|T_{11}|^2+|T_{1-1}|^2},\hspace*{5mm}
r_{1-1}^{04} \simeq \frac{\strut {\rm Re}(T_{11}T_{1-1}^*)}{\strut
    |T_{01}|^2+|T_{11}|^2+|T_{1-1}|^2}, \nonumber \\[3mm]
{\rm Re}(r_{10}^{04}) \simeq \frac{1}{2} \left [ \frac
{\strut {\rm Re}(T_{11}T_{01}^*) + {\rm Re}(T_{1-1}T_{0-1}^*)}{\strut |T_{01}|^2+|T_{11}|^2+|T_{1-1}|^2} 
\right].
\end{eqnarray}

\begin{figure}[htbp]
\begin{center}
\epsfig{file=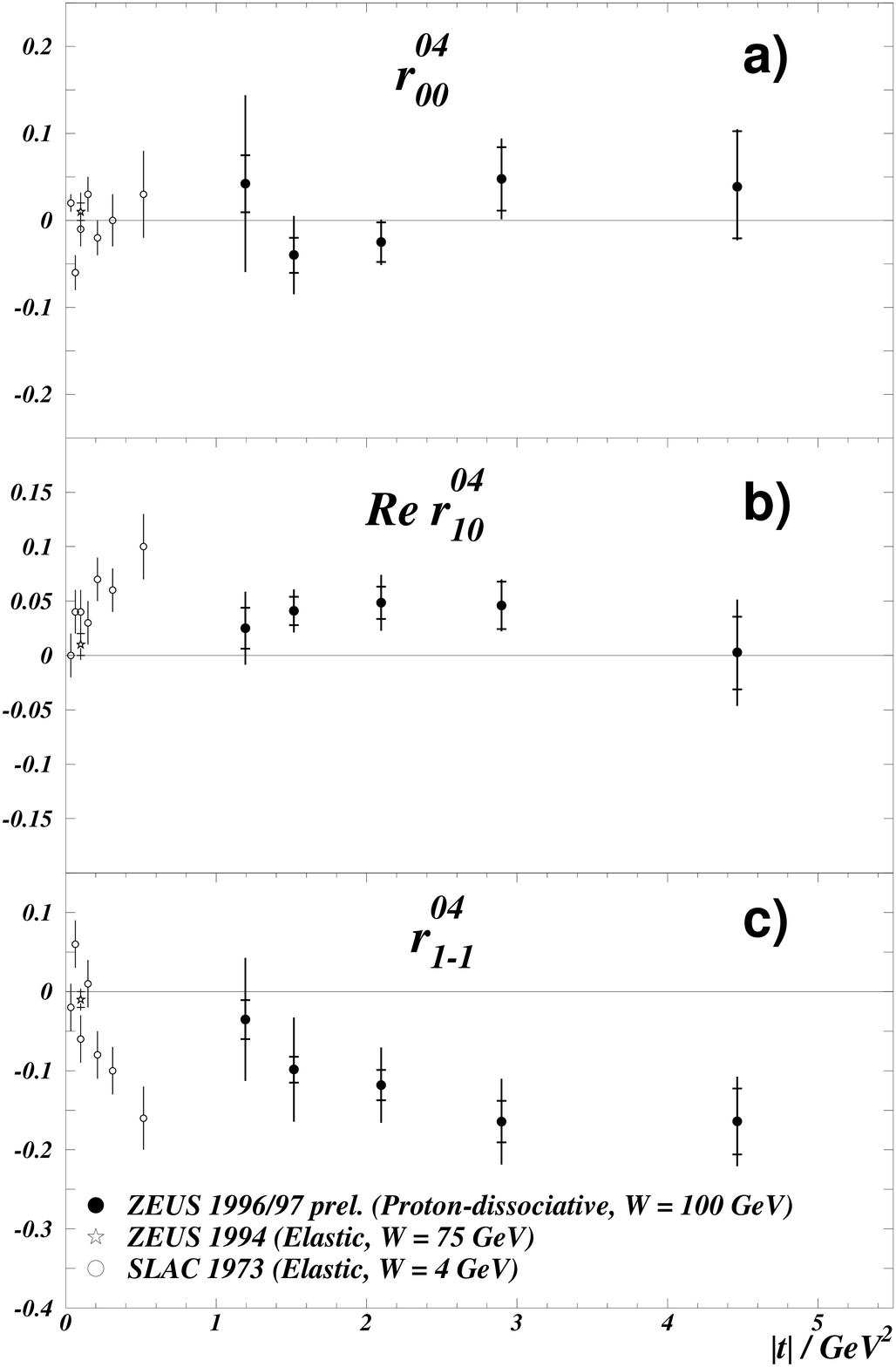, width=0.62\textwidth, bbllx=0pt, bblly=0pt, bburx=800pt, bbury=1227pt, clip=}
\caption{
Recent measurements by the ZEUS  
collaboration~\protect\cite{hep99_06_005,eps99_499} 
of the combinations of matrix elements  (a)~$r^{04}_{00}$, (b)~Re~$r^{04}_{10}$, (c)~$r^{04}_{1-1}$ for the diffractive  photoproduction of pion pairs. 
The inner error bars represent the statistical uncertainties;
the outer error bars show the statistical and systematic uncertainties added
in quadrature. See text for full description}
\label{fig:rij}
\end{center}
\end{figure}

Figure~\ref{fig:rij}
shows the results for the combinations
of matrix elements obtained from a least-squares minimization procedure
in which they served as fit parameters.
The inner error bars represent the statistical
uncertainty; the outer error bars show the quadratic sum of
statistical and systematic uncertainties. 
The systematic
uncertainties are dominated by the uncertainty in the acceptance corrections.
The dipion mass range was restricted to the region
\mbox{$0.45\hspace*{-2pt}<\hspace*{-2pt}M_{\pi\pi}\hspace*{-2pt}<\hspace*{-2pt}1.1~{\gev}$}.
The results are compared to the results at lower {\abst} for the exclusive
reaction obtained with 9 {\gev}
photons from a backscattered laser beam  incident on a hydrogen bubble 
chamber at SLAC.\hspace*{-0.7ex}~\cite{pr_7_3150} Also shown are the ZEUS 1994 results for exclusive {\rhoz}
photoproduction at low {\abst}.\hspace*{-0.7ex}~\cite{epj_2_247} 
The parameter $r^{04}_{00}$
is consistent with zero over the entire range in {\abst},
showing no evidence for the production of {\rhoz} mesons in helicity
state 0. 
The combination \mbox{Re $r^{04}_{10}$}, which is predominantly sensitive to the interference between
the helicity-conserving amplitude and the single-flip amplitude, shows clear
evidence for a small single-flip contribution in both the SLAC data and the
high-{\abst} ZEUS results. A clear indication of a sizeable double-flip contribution is
shown by the measurements of $r^{04}_{1-1}$ at high {\abst}, as was seen in
the SLAC results at lower {\abst} and lower energy. 

In order to estimate the effect on the angular distributions of a 
hypothesized dipion background to {\rhoz} decay, the decay-angle 
analysis was repeated for restricted dipion mass ranges above
(\mbox{$0.77<M_{\pi\pi}<1.0~{\gev}$}) and below
(\mbox{$0.6<M_{\pi\pi}<0.77~{\gev}$}) the nominal 
value for the {\rhoz} mass. 
The observed value for each of the combinations of 
matrix elements was found to depend
significantly
on the mass range chosen, though the above conclusions concerning the lack of
longitudinal polarization and the evidence for a double-flip contribution
remain unchanged. This
dependence on dipion invariant mass may suggest the presence of a non-resonant background. The extraction of the spin-density
matrix elements for the {\rhoz} meson alone 
from these dipion angular distributions thus
awaits the understanding of this dependence.

The complexity of the helicity structure in pion-pair photoproduction
is thus shown to persist at high values of {\abst},
where the hardness of the {\abst} spectrum and the value of 
the $\phi$/{\rhoz} ratio
encourage attempts to describe the semi-exclusive electroproduction of vector mesons in the 
framework of perturbative QCD. 
\section{Concluding Remarks}
The selection of results on diffractive vector-meson production from
the HERA collider experiments presented above exemplifies
ongoing measurement programs 
undertaken by the H1 and ZEUS collaborations. The luminosity upgrade
program scheduled for the year 2000 is expected to provide an increase
in instantaneous luminosity of roughly a factor of five. This increase
is of particular importance for investigations such as those covered above
in the context of this workshop: $\Upsilon$ photoproduction, complete
helicity analyses of vector-meson electroproduction, and 
semi-exclusive vector-meson photoproduction at high momentum transfer,
since their accuracy is statistically limited. The future operation
of HERA is therefore certain to continue to provide new results on
exclusive and semi-exclusive processes at hard scales, ensuring
a rigorous series of tests for new applications of quantum chromodynamics.
\section*{Acknowledgments}
The author thanks the organizers  for their hospitality during the workshop.
Figure~\ref{fig:upsrot} was provided by P.~Merkel. This work is supported by the Federal Ministry of Education and Research of
Germany.

\section*{References}
\bibliographystyle{zeusstylem}
\bibliography{zeuspubs,h1pubs,otherpubs}

\end{document}